\documentstyle[prb,aps]{revtex}

\begin{document}
\twocolumn[\hsize\textwidth\columnwidth\hsize\csname 
@twocolumnfalse\endcsname

\title{Influence of nonlocal electrodynamics on the anisotropic vortex pinning in $YNi_2B_2C$}
\author{A.V. Silhanek$^1$, J.R. Thompson$^2$, L. Civale$^1$, D.McK. Paul$^3$ and C.V. Tomy$^3$}
\address{$^1$Comisi\'{o}n Nacional de Energ\'{\i}a At\'{o}mica-Centro At\'{o}mico Bariloche\\ and Instituto Balseiro, 8400 Bariloche, Argentina.\\$^2$Oak Ridge National Laboratory, Oak Ridge, Tennessee 37831-6061.\\Department of Physics, University of Tennessee, Knoxville, Tennessee 37996-1200.\\$^3$Department of Physics, University of Warwick, Coventry, CV4 7AL, United Kingdom}

\date{August 23, 2000}
\maketitle

\begin{abstract}
We have studied the pinning force density $F_p$ of $YNi_2B_2C$ superconductors for various field orientations. We observe anisotropies both between the c-axis and the basal plane and within the plane, that cannot be explained by usual mass anisotropy. For magnetic field ${\bf H}\parallel c$, the reorientation structural transition in the vortex lattice due to nonlocality, which occurs at a field $H_1 \sim 1kOe$, manifests itself as a kink in $F_p(H)$. When ${\bf H} \bot c$, $F_p$ is much larger and has a quite different $H$ dependence, indicating that other pinning mechanisms are present. In this case the signature of nonlocal effects is the presence of a fourfold periodicity of $F_p$ within the basal plane.

\end{abstract}

\pacs{PACS 74.60.Ge; 74.60.Dh; 74.60.Jg}
\vskip1pc] \narrowtext

After Kogan and co-workers\cite{kogan96} showed in 1996 that nonlocal electrodynamics effects are relevant in high-$\kappa$ superconductors in a broad H-T domain, several works demonstrated that nonlocality give rise to unusual properties of the vortex matter. Those studies were done mainly on the family of compounds $\it{R}Ni_2B_2C$, where $\it{R} = Lu, Y, Tm, Er, Ho, Dy$. These borocarbides are very well suited to study nonlocal effects due to their intermediate $\kappa$ values ($\sim$ 10 - 20) that result in a wide field range where the extended London description holds, their high transition temperature $T_c$ and the possibility to fabricate clean single crystals with large electronic mean free path. Deviations of the equilibrium magnetization from the local London prediction,\cite{song99} structural phase transitions separating a variety of exotic rhombic and square vortex lattices (FLL),\cite{yethiraj97,kogan97a,eskildsen97b,dewilde97,yethiraj98,mckpaul98,gammel99,sakata00} and fourfold anisotropies in the basal plane of these tetragonal materials\cite{civale99,kogan99} provide strong support for the nonlocal scenario. 

Although the consequences of nonlocality on the $\it {equilibrium}$ properties of the superconducting vortex matter is by now convincingly established, very little is known about its effects on the $\it {nonequilibrium}$ vortex response. For both $YNi_2B_2C$ and $LuNi_2B_2C$ Eskildsen et al.\cite{eskildsen97a} have shown that, for ${\bf H}\parallel c$-axis and well above the field $H_2 \sim 1$kOe marking a transition from a rhombic to a square vortex lattice, the pinning properties sharply disagree with the expectations of the Larkin-Ovchinnikov collective pinning theory for a triangular lattice. The authors associated the discrepancy to the different (and to large extent unknown) elastic properties of the square lattice. However, neither the irreversible response in other field orientations, where the lattice geometry is different, nor the influence of the lattice transitions themselves on vortex pinning have been experimentally explored until now. It is clear that vortex pinning, which involves distortions from equilibrium vortex configurations, should be affected by the symmetry changes in the lattice.

In this work we report a study of nonlocal effects on the vortex pinning of a single crystal of $YNi_2B_2C$. For ${\bf H}\parallel c$-axis the field dependence of the pinning force density $F_p$ exhibits a distinct kink at a field $H^*$ which coincides with the first order reorientation transition in the vortex lattice. The angular dependence of $H^*$ (for $\bf{H}$ inclined up to $30^{\circ}$ from $c$) 
provide further support for the association of the kink with the FLL transition. For $\bf{H}$ in the ab-plane, $F_p$ is an order of magnitude larger and the field dependence is quite different. We discuss the possible origin of this anisotropy. Finally, we observe fourfold oscillations in the angular dependence of $F_p$ in the basal plane which also arise from nonlocal effects. 

The single crystal measured in the present study, with a mass of 17 mg, dimensions $\sim 2 \times 2.5 \times 0.5 mm$, and critical temperature $T_c$=14.5K, is the same one used in previous works.\cite{song99,civale99}. Electrical resistivity data yielded a residual resistance ratio of 10 and an estimate of the electronic mean free path ${\it l} \sim 300 \AA$. Magnetic measurements were carried out in two Quantum Design SQUID magnetometers. All data were taken in a MPMS-7 ($H \leq 7T$) except the angular dependence in the basal plane, which was studied in a MPMS-5S ($H \leq 5T$) using a home made rotating system.\cite{silhanek99} In the superconducting mixed state, isothermal magnetization loops were recorded. The hysteresis width $\Delta M (H)$ was used to calculate the critical current density $J_c$ from the Bean's critical state model.

Figure 1(a) shows the pinning force density $\bf{F}_p=\bf{J_c} \times \bf{B}$ as a function of ${\bf H}\parallel c$-axis at several temperatures. At $T=3K$, after an initial increase at very low fields (not shown), $F_p (H)$ reaches a maximum at a remarkably low reduced field ($H/H_{c2}<0.01$), and then decreases approximately linearly in $log(H)$. At a field $H^* \sim 1$kOe a clear kink is observed. Above this field $F_p$ is still linear in $log(H)$, but the slope $dF_p/dlogH$ has a smaller value. At a higher field $H_b \sim 10$kOe a small bump is visible. As discuss below, we associate $H_b$ with the peak observed by Eskildsen et al.\cite{eskildsen97a} in $LuNi_2B_2C$. Finally, close to $H_{c2}$ we observe the usual peak effect attributed to the softening of the flux line lattice.

In the framework of Kogan's model\cite{kogan97a} two structural transitions in the vortex lattice occur in $YNi_2B_2C$, namely a first order reorientation transition between two rhombic lattices at a field $H_1$ and a second order transition from the rhombic to a square lattice at a higher field $H_2$. According to a recent analysis by Knigavko et al.\cite{knigavko} the reorientation at $H_1$ really consists of two second order transitions, but they take place in such a narrow field range that may be hard to resolve them experimentally. Small angle neutron scattering (SANS) experiments\cite{mckpaul98} (performed on a crystal of the same batch as ours) confirmed Kogan's predictions. For ${\bf H} \parallel c$, a jump in the apical angle $\beta$ of the rhombic vortex lattice (discontinuous within the resolution) occurs at $H_1 \sim 1$ to $1.25kOe$, and the lattice becomes square ($\beta = 90^{\circ}$) at $H_2 \sim 1.25$ to $1.5kOe$. The coincidence of our $H^*$ with this field range make it tempting to associate the jump in $dF_p/dlogH$ with one of these transitions. However, the small difference between $H_1$ and $H_2$ make it difficult to determine which one of them generates the kink. 

To solve this question we need to modify the experimental conditions in such a way that $H_1$ and $H_2$ change and split apart. The simplest way is to change $T$. Fig. 1(a) shows that as $T$ increases $H^*$ remains almost constant, while the slope change progressively washes out. This last observation is consistent with the association of $H^*$ with nonlocality, as those effects tend to disappear as $T$ approaches $T_c$. On the other hand, theoretical expectations\cite{kogan97a} as well as some qualitative experimental evidence\cite{eskildsen97b} suggests that $H_2$ increases with $T$ well below $T_c$, in contrast with the observed behavior of $H^*$. This seems to indicate that the kink is not related with $H_2$. 

To our knowledge there are no experimental data on the temperature dependence of $H_1$. The only theoretical hint comes from the recent study of Knigavko et al.\cite{knigavko}. It is shown there that $H_1$ in units of the scaling field $\Phi_0 / (2\pi \lambda)^2$ is a decreasing function of the strength of the nonlocal effects (which is parametrized by two averages over the Fermi surface, $n$ and $d$), and consequently must increase with $T$. Disregarding a weak temperature dependence in $\kappa$, the scaling field is proportional to $H_{c2}$, which is easily accessed experimentally. This suggests to analyze $F_p$ as a function of the reduced field $h=H/H_{c2}$.

In Figure 2 we replotted the data of Fig. 1(a) as $f_p=F_p/F_p^{peak}$ vs $h$. Here $F_p^{peak}$ is the maximum of $F_p$ at the peak close to $H_{c2}$, which occurs at the same reduced field $h_p=0.8$ for all T. The overlap of the data at different $T$ in the $h$ range of the peak-effect means that the pinning mechanism responsible for it satisfies a scaling law $F_p(H,T)=F_p^{peak}(T)f_p(h)$, as seen in many systems.\cite{kramer73} On the other hand, the fact that at lower $h$ the $f_p(h)$ data at various $T$ do not collapse on a single curve indicates that the dominant pinning mechanism(s) in that range is(are) different from that originating the peak-effect. As already mentioned, we identify the bump at intermediate field with the peak reported by Eskildsen et al.\cite{eskildsen97a} The fact that this bump occurs at the same $h_b \sim 0.18$ for all $T$ is consistent with the interpretation of this feature as resulting from the stiffening of the vortex matter due to the increase of the shear modulus $C_{66}$, as this elastic constant depends on field only through $h$.\cite{eskildsen97a} 

In contrast to $h_p$ and $h_b$, Fig. 2 shows that $h^*=H^*/H_{c2}$ is temperature dependent. This lack of scaling is expected for any feature arising from nonlocal effects, as they involve a new characteristic field $H_0 \propto n^{-2}$, whose $T$ dependence is quite different from that of $H_{c2}$. In particular, the fact that $h^*$ increases with $T$ is consistent with the theoretical prediction for $h_1=H_1/H_{c2}$ as discussed above. So, from the temperature dependence of $H^*$ we conclude that it is probably associated with $H_1$.

Further evidence that $H^*$ is a signature of the reorientation transition at $H_1$ arises from the angular dependence. SANS results show\cite{mckpaul98} that $H_2$ is very sensitive to the field orientation, while $H_1$ is much less so. When $\bf {H}$ is tilted by an angle $\Theta = 10^{\circ}$ from the c-axis, $H_1$ remains unchanged within the resolution, while $H_2$ shifts up significantly. Extrapolation of the data (the maximum measured field was $2.5kOe$) gives $H_2 \sim 2.7kOe$. For $\Theta = 30^{\circ}$ only four SANS data points are available. Now $\beta$ increases only weakly with $H$, and extrapolation suggests that $H_2$ is well above $4kOe$. The inset of Fig. 2 shows $F_p$ vs $log(H)$ at $T=3$ and $5K$, for $\Theta = 0^{\circ}$, $10^{\circ}$ and $30^{\circ}$. The $10^{\circ}$ curves are almost identical to ${\bf  H} \parallel c$ and the kink is clearly visible at the same $H^*$, indicating that it is related to $H_1$ rather than to $H_2$. In the $\Theta = 30^{\circ}$ curves $H^*$ is slightly shifted up, while according to the SANS data $H_1$ has either disappeared or shifted down to $H < 1 kOe$. However the SANS evidence in this case is inconclusive, as it is based on a single data point and it may be affected by phase coexistence.

The very low $J_c$ for ${\bf H} \parallel c$ indicates that pinning correlation volumes are large, as confirmed by SANS results.\cite{eskildsen97a} Thus, the elastic properties of the vortex lattice must play a key role in the pinning. While the tilt modulus $C_{44}=BH/4\pi$ should be rather independent of the lattice details, it is easy to see that $C_{66}$ depends on $\beta$. As $\beta$ overgoes a discontinuous jump at $H_1$, it is natural to ascribe the kink to an abrupt change in $C_{66}$. It has been argued\cite{eskildsen97a} that below $H_b$ the collective pinning description is valid in $YNi_2B_2C$. In the simplest scenario this implies that $F_p \propto C_{44}^{-1}C_{66}^{-2}$. Using this expression and the experimental $F_p(H)$ we computed $C_{66}(H)$, and indeed found a kink at $H^*$ as expected. The surprising result is that $C_{66}$ decreases monotonically with H both below and above $H^*$, in contradiction with the standard behavior. It must be noted, however, that the shear properties for $\beta \neq 60^{\circ}$ are anisotropic, thus $C_{66}$ has more than one component and the expression for $F_p$ should be modified.

We now turn to the pinning properties for ${\bf H} \bot c$. Figure 1(b) shows $F_p(H)$ for ${\bf H}\parallel [100]$ and ${\bf H} \parallel [110]$ at various $T$. The behavior in these orientations is very different from the ${\bf H} \parallel c$ case. At $T=3K$ a broad maximum occurs at fields $H_{max} \sim 8kOe$ for ${\bf H} \parallel [100]$ and slightly less for ${\bf H} \parallel [110]$. Over most of the field range $F_p$ is much larger than for ${\bf H} \parallel c$, the maximum anisotropy being a factor of $\sim 20$ at $H_{max}$. As $T$ increases, $H_{max}$ decreases while the out-of-plane anisotropy in $F_p$ remains approximately constant. Another feature visible in Fig. 1(b) is the in-plane anisotropy, $F_p[100]>F_p[110]$. The difference is small, but clear and systematic, and holds at all $H$ below the peak effect. This in-plane anisotropy smoothly decreases with $T$.

The large difference in $F_p$ between the c-axis and the basal plane is surprising, considering that the mass anisotropy\cite{kogan99} is very small ($< 10 \%$) as confirmed by the $H_{c2}$ values in Figs. 1(a) and (b). This rules out explanations based on the anisotropic scaling frequently used in high $T_c$ superconductors.\cite{blatter92} And of course mass anisotropy cannot account for any angular dependence in the basal plane of this tetragonal compound. We will now discuss the possible sources of both the out-of-plane and in-plane anisotropy.

From the analysis of the normal state magnetization $M_{ns}$ we found\cite{civale99} that this crystal contains a small amount of magnetic impurities, about $\sim 0.1\%$ of a magnetic rare earth substituting for Y. The moments lay in the basal plane and produce a Curie tail that, for values of the scaling variable $H/T$ up to $\sim 1kOe/K$, is isotropic within the plane. For higher $H/T$ a fourfold in-plane anisotropy develops. This is shown in the inset of Fig. 3, where $\delta M_{ns}=M_{ns}[110]-M_{ns}[100]$ is plotted as a function of $H/T$. These localized moments are a potential source of anisotropic vortex pinning. However, although we cannot completely rule out this possibility, several observations suggest that this is not the case.

In the first place, as the alignment of the moments increases with $H$, the directional pinning should become more efficient as the field increases and thus the out-of-plane anisotropy in $F_p$ should increase monotonically with $H$. This is not in agreement with the observations. At $T=3K$ the anisotropy factor is already large ($\sim 6$) at $H \sim 1kOe$, then it grows with $H$, maximizes at $H \sim H_{max}$ and decreases again for higher fields. In particular, $F_p$ is almost isotropic (within $\sim 10 \%$) in the peak effect.

Another clue comes from the in-plane anisotropy. As $M_{ns}$ is isotropic within the plane for $H/T<1kOe/K$ (see inset of Fig. 3), we would expect $F_p$ to be also isotropic in the plane below that threshold. This is again contrary to the results: Fig 1(b) shows that at all temperatures $3K<T<9K$ the in-plane anisotropy in $F_p$ is already visible at fields well below the $H/T$ threshold. 

The lower panel of Figure 3 shows the in-plane angular dependence of the pinning force density, $F_p(\varphi)$, at $T=7K$ for several $H$, and the upper panel shows $M_{ns}(\varphi)$ for $H=45kOe$. This figure clearly demonstrates that $F_p$ exhibits a fourfold anisotropy for all $H>2kOe$, while at this temperature the threshold for the anisotropy in $M_{ns}$ is $\sim 7kOe$. The behavior of $F_p(\varphi)$ is rather complex: in addition to the main peaks at [100] and [010], secondary maxima are visible at [110] and (1$\bar{1}$0). Further evidence that the localized moments are not the source of anisotropic pinning is that the fourfold oscillations in $M_{ns}$ and $F_p$ have opposite sign, as is evident by comparing the lower and upper panels. The presence of four- and higher-fold oscillations in Fig. 3, as well as the nontrivial $T$ and $H$ dependencies in Fig. 1(b), rule out the possibility that the in-plane anisotropy could be an artifact due to a slight misalignment between ${\bf H}$ and the basal plane. Such misalignment would only produce two-fold oscillations, which are not observed.

Once the possibility that the in-plane anisotropy in $F_p$ is due to the magnetic impurities has been ruled out, we must consider other alternatives. While it is very unlikely that any type of crystallographic defects could account for the complex in-plane variations in $F_p$, nonlocal effects can provide a natural explanation for them. Due to nonlocality, the geometry of the vortex lattice depends on the orientation within the plane. We can again argue (as in the ${\bf H} \parallel c$ case) that $C_{66}$ depends on the lattice geometry, and that such dependence must be reflected in $F_p$. Interestingly $F_p$ has local maxima at the high symmetry crystallographic orientations [100] and [110], where $C_{66}$ is expected to exhibit local minima. The fact that the in-plane anisotropy disappears as $T$ approaches $T_c$ provide further support to this interpretation. 

The origin of the out-of-plane anisotropy in $F_p$ is less clear. On one hand, it seems too large to be ascribed to nonlocal effects. On the other hand, a simpler explanation for it could be the presence of some still unidentified anisotropic pinning centers, such as planar defects. Clearly, the unexpected complexity of the angular dependence of $F_p$ involves other mechanisms besides nonlocality, and further studies are necessary to sort out the various sources of anisotropic pinning in this material.

In conclusion, $F_p$ in $YNi_2B_2C$ exhibits a rich anisotropic behavior that sharply contrasts with its small mass anisotropy. Nonlocal electrodynamics influences pinning via the unusual behavior of the shear modulus in non-hexagonal lattices, either through continuous variations (fourfold basal plane anisotropy) or abrupt jumps (the kink at $H_1$). A complete understanding of the vortex irreversible response will require a deeper knowledge of the elastic properties of non-hexagonal lattices and the extension of the pinning models to those structures. The origin of the large out-of plane pinning anisotropy, which almost certainly does not originate on nonlocality, also deserves further investigation.

Work partially supported by ANPCyT, Argentina, PICT 97 No.01120, and CONICET PIP 4207. A.V.S. is member of CONICET. Research sponsored by the U.S. Department of Energy under contract DE-AC05-00OR22725 with the Oak Ridge National Laboratory, managed by UT-Battelle, LLC. We thank Hyun Jeong Kim for her valuable help in conducting experiments. We acknowledge useful discussions with D.K. Christen, M. Yethiraj and H.R. Kerchner. 

\section{References}

\bibliographystyle{prsty}

Fig.1: Pinning force density $F_p$ vs $H$ for $YNi_2B_2C$ single crystal with (a) $\bf{H} \parallel c$-axis and (b) $\bf{H} \parallel$ [100] (open symbols) and $\bf{H} \parallel$ [110] (solid symbols) at several temperatures.

Fig.2: Reduced pinning force density $F_p/F_p^{max}$ vs $H/H_{c2}$ at several temperatures with $\bf{H} \parallel c$-axis. The inset shows $F_p$ versus $H$ for ${\bf H}\parallel c$ ($\Theta = 0^\circ$) and tilted $10^\circ$ and $30^\circ$ from $c$.

Fig.3: Upper panel: the normal state magnetization $M_{ns}$ at $T=7K$ and $H=45 kOe$, as a function of the angle $\varphi$ between the applied field $\bf H$ (contained in the $\it {ab}$ basal plane) and the $\it{a}$ axis. Lower panel: pinning force density $F_p (\varphi)$ at $T=7K$ and several fields. Inset: oscillation amplitude of the in-plane normal state magnetization, $\delta M_{ns}=M_{ns}[110]-M_{ns}[100]$, vs $H/T$.

\end{document}